\documentclass[10pt]{amsart}
 
\usepackage{amsthm,amsmath,amssymb}
\usepackage{graphicx}
\usepackage[comma, sort&compress]{natbib}

\usepackage{paralist}
\RequirePackage{hyperref}

\newtheorem{theorem}{Theorem}

\theoremstyle{definition} 

\theoremstyle{remark}

\numberwithin{equation}{section}
\numberwithin{theorem}{section}
\numberwithin{example}{section}
\numberwithin{definition}{section}

\numberwithin{figure}{section}

\title[]{Order-invariant prior specification in Bayesian factor analysis}

\author[D.~Leung]{Dennis Leung} 
\address{Department of Statistics, University of Washington, Seattle,
  WA, U.S.A.}
\email{dmhleung@uw.edu}

\author[M.~Drton]{Mathias Drton} 
\address{Department of Statistics, University of Washington, Seattle,
  WA, U.S.A.}
\email{md5@uw.edu}

\begin{document}

\begin{abstract}
  In (exploratory) factor analysis, the loading matrix is identified
  only up to orthogonal rotation.  For identifiability, one thus often
  takes the loading matrix to be lower triangular with positive
  diagonal entries.  In Bayesian inference, a standard practice is
  then to specify a prior under which the loadings are independent,
  the off-diagonal loadings are normally distributed, and the diagonal
  loadings follow a truncated normal distribution.  This prior
  specification, however, depends in an important way on how the
  variables and associated rows of the loading matrix are ordered.  We
  show how a minor modification of the approach allows one to compute
  with the identifiable lower triangular loading matrix but maintain
  invariance properties under reordering of the variables.
\end{abstract}



\maketitle

\section{Introduction}
\label{sec:introduction}

Let $y$ be an $m$-vector of observed random variables, which for
simplicity we take to be centered.  Let $f\sim N_k(0,I_k)$ be a standard
normal $k$-vector of latent factors, with $k\le m$.  The factor
analysis model postulates that
\begin{equation} 
  \label{vecModel}
  {y} = {\beta}{f} + {\varepsilon}, 
\end{equation}
where $\beta=(\beta_{ij})\in\mathbb{R}^{m\times k}$ is an unknown
loading matrix, and $\varepsilon\sim N_m(0,\Omega)$ is an $m$-vector
of normally distributed error terms that are independent of $f$.  The
error terms are assumed to be mutually independent with
$\Omega=\text{diag}(\omega^2_1, \dots, \omega^2_m)$ comprising $m$
unknown positive variances that are also known as uniquenesses.  This
model with an unrestricted $m\times k$ loading matrix $\beta$ is
sometimes referred to as exploratory factor analysis---in contrast to
confirmatory factor analysis, which refers to situations in which some
collection of entries of $\beta$ is modeled as zero.

Integrating out the latent factors $f$ in~(\ref{vecModel}), the
observed random vector $y$ is seen to follow a centered multivariate
normal distribution with covariance matrix
\begin{equation}
  \label{eq:y:sigma}
  \Sigma = \Omega + \beta\beta'.
\end{equation}
As discussed in detail in \cite{MR0084943}, $\Sigma$ determines the
unrestricted loading matrix $\beta$ only up to orthogonal rotation.
Indeed, $\beta\beta'=\beta QQ'\beta'$ for any $k\times k$ orthogonal
matrix $Q$.  More details on factor analysis can be found, for
instance, in
\cite{MR2849614}, \cite{MR2299716},
and \cite{MR2604888}.

In this paper, we are concerned with Bayesian inference in
(exploratory) factor analysis.  In Bayesian computation, it is
convenient to impose an identifiability constraint on the loading
matrix $\beta$.  A common choice is to restrict $\beta$ to be lower
triangular with nonnegative diagonal entries, that is, $\beta_{ij}=0$
for $1\le i<j\le k$ and $\beta_{ii}\ge 0$ for $1\le i\le k$
\citep{geweke:1996,aguilar:2000,MR2036762}.  Under these constraints,
a full rank matrix $\beta$ is uniquely determined by $\beta\beta'$.
In the papers just referenced and also the software implementation
provided by \cite{mcmcpack}, a default prior on the lower triangular
loading matrix has all its non-zero entries independent with
\begin{align}
  \label{eq:beta:prior:standard}
  \beta_{ij} &\sim 
  \begin{cases}
    \mathit{TN}(0, C_0) &\text{ if } i=j,\\
    N(0, C_0) &\text{ if } i>j.
  \end{cases}
\end{align}
Here, $\mathit{TN}(0,C_0)$ denotes a truncated normal distribution on
$(0,\infty)$, i.e., the conditional distribution of $X$ given $X>0$
for $X\sim N(0,C_0)$.  The variance $C_0>0$ is a hyperparameter.  The
prior distribution for the uniquenesses has
$\omega_1^2,\dots,\omega_m^2$ independent of $\beta$ and also mutually
independent with Inverse Gamma distribution,
\begin{align}
  \label{eq:IGprior}
  \omega_i^2&\sim \mathit{IG}(\nu/2, \nu s^2/2)
\end{align}
for hyperparameters $\nu,s>0$.  Equivalently, $\nu s^2/\omega_i^2$ is
chi-square distributed with $\nu$ degrees of freedom; compare
Eqn.~(26) in
\cite{geweke:1996}.

As discussed in \citet[Sect.~6]{MR2036762}, the prior specification
in~(\ref{eq:beta:prior:standard}) is such that the induced prior on
$\beta\beta'$ and the covariance matrix $\Sigma$ in
(\ref{eq:y:sigma}) depends on the way the variables and the associated
rows of the loading matrix $\beta$ are ordered.  Indeed, a priori,
\begin{equation}
  \label{eq:betabeta:ii:chisquare}
  (\beta\beta')_{ii}/C_0=\sum_{j=1}^k
  \beta_{ij}^2/C_0=\sum_{j=1}^{\min\{i,k\}}
  \beta_{ij}^2/C_0
\end{equation}
follows a chi-square distribution with degrees of freedom
$\min\{i,k\}$.  Consequently, the implied prior and also the
posterior distribution for the covariance matrix $\Sigma$ is not
invariant under permutations of the variables.

In this paper we propose a modification of the prior distribution for
$\beta$ that maintains the convenience of computing with an
identifiable lower triangular loading matrix all the while making the
prior distributions of $\beta\beta'$ and $\Sigma$ invariant under
reordering of the variables.  Our proposal, described in
Section~\ref{sec:order-invariant}, merely changes the prior
distributions of the diagonal entries $\beta_{ii}$
in~(\ref{eq:beta:prior:standard}), which will be taken from a slightly
more general family than the truncated normal.  The details of a Gibbs
sampler to draw from the resulting posterior are given in
Section~\ref{sec:gibbs}.  We conclude with numerical examples and a
discussion in Sections~\ref{sec:numerics} and~\ref{sec:conclusion},
respectively.

\section{Order-invariant prior distribution}
\label{sec:order-invariant}

Without any identifiability constraints, the loading matrix $\beta$
takes its values in all of $\mathbb{R}^{m\times k}$.  A natural
default prior would then be to take all entries $\beta_{ij}$,
$i=1,\dots,m$ and $j=1,\dots, k$, to be independent $N(0,C_0)$ random
variables; we write $\beta\sim N_{m\times k}(0,C_0I_m\otimes I_k)$.
The spherical normal distribution $N_{m\times k}(0,C_0I_m\otimes I_k)$
is clearly invariant under permutation of the rows of the matrix.
Hence, the induced prior distribution of $\beta\beta'$ and of the
covariance matrix $\Sigma$ from~(\ref{eq:y:sigma}) is invariant under
simultaneous permutation of rows and columns.

Working with the prior just described comes at the cost of losing the
identifiability of $\beta$.  However, this can be overcome as follows.
Assuming that $m \geq k$, any $m \times k$ matrix $\beta$
with linearly independent columns can be uniquely decomposed as $\beta
= LQ$, where $L$ is an $m \times k$ lower triangular matrix with
positive diagonal, and $Q$ is a $k \times k$
orthogonal matrix.  We may then use the implied distribution of the
lower triangular matrix $L$ as a prior on the loading matrix.  The
following theorem about the joint distribution of $L$ and $Q$ is
adapted from Theorem $2.1.13$ in \cite{MR652932}.

\begin{theorem} 
  \label{LQthm} 
  Let $\beta=LQ$ be the LQ decomposition of the $m\times k$ random
  matrix $\beta\sim N_{m\times k}(0,C_0I_m\otimes I_k)$, where $m\ge k$.  Then the
  lower triangular matrix $L$ and the orthogonal matrix $Q$ are
  independent, the distribution of $Q$ is the normalized Haar measure,
  and the distribution of $L=(L_{ij})$ has joint density
  proportional to
  \begin{equation}
    \label{Lprior}
    \prod_{i=1}^m\prod_{j=1}^{\min\{i,k\}} \exp\left\{-\frac{1}{2C_0}L_{ij}^2 \right\}\times
    \prod_{i = 1}^k L_{ii}^{k - i} \mathbf{1}_{\{L_{ii}>0 \}}
  \end{equation}
  with respect to the Lebesgue measure on the space of $m \times k$
  lower triangular matrices.
\end{theorem}

The joint distribution for the entries of $L=(L_{ij})$ given
by~(\ref{Lprior}) has the entries $L_{ij}$, $i\ge j$, independent with
$L_{ij}\sim N(0,C_0)$ if $i>j$ and $L_{ii}$ following the distribution
with density proportional to
\begin{equation}
  \label{eq:new-distribution}
  x^{k-i}\exp\left\{-\frac{1}{2C_0}x^2\right\}, \quad x>0.
\end{equation}
Note that $L_{kk}\sim\mathit{TN}(0,C_0)$.  The joint distribution for
a lower triangular matrix in~(\ref{Lprior}) thus differs from that
given by~(\ref{eq:beta:prior:standard}) only in the coordinates
$L_{ii}$ for $1\le i\le k-1$, which are no longer truncated normal.

Assume as in~(\ref{eq:IGprior}) that $\Omega$ and $\beta$ are
independent a priori.  Then since $Q$ is independent of $L$, and
\[
\Sigma=\Omega+\beta\beta'=\Omega+LQQ'L' = \Omega+LL'
\]
does not depend on $Q$, the tuple $(y,\Omega,L,\Sigma)$ is independent
of $Q$.  Hence, $(\Omega,L,\Sigma)$ is also independent of $Q$ a
posteriori (i.e., conditional on $y$).  Our proposal is now simply to
keep with the standard identifiability constraint that has the loading
matrix $\beta$ lower triangular with nonnegative diagonal entries but to
use the distribution given by~(\ref{Lprior}) instead
of~(\ref{eq:beta:prior:standard}) for this lower triangular loading
matrix.  Concerning the remaining parts of the prior specification, we
continue to assume independence of $\beta$ and $\Omega$, and we stick
with the choice from~(\ref{eq:IGprior}) for the prior on the
uniquenesses.  This proposed prior has then the property that the
distributions of $\beta\beta'$ and the covariance matrix $\Sigma$ are
invariant under reordering of the variables (i.e., matrix rows and
columns), both a priori and a posteriori.

\section{Gibbs sampler}
\label{sec:gibbs}

Consider now an actual inferential setting in which we observe a
sample $y_1,\dots,y_n$ that comprises $n$ independent random vectors
drawn from a distribution in the $k$-factor model.  Let $Y$ be the
$n\times m$ matrix with the vectors $y_1,\dots,y_n$ as rows.  Let $F$
be an associated $n\times k$ matrix whose rows $f_1, \dots,f_n$ are
independent vectors of latent factors.  The factor analysis model
dictates that
\begin{equation} 
  \label{matModel}
  {Y} = {F} {\beta}' + E, 
\end{equation}
where $E= (\varepsilon_1, \dots, \varepsilon_n)'$ is an $n\times m$
matrix of stochastic errors.  The pairs $(f_t, \varepsilon_t)$ for
$1\le t\le n$ are independent, and in each pair $f_t\sim N_k(0,I_k)$
and $\varepsilon_t\sim N_m(0,\Omega)$ are independent as well.  The
unknown parameters are comprised in the matrices $\Omega =
\text{diag}(\omega^2_1, \dots, \omega^2_m)$ and $\beta =
(\beta_{ij})\in\mathbb{R}^{m\times k}$, where the latter is restricted to
be lower triangular with nonnegative diagonal.

We now adopt the prior distribution on $\beta$ and $\Omega$ given
by~(\ref{Lprior}) and~(\ref{eq:IGprior}), and derive the full
conditionals needed for a Gibbs sampler that generates draws from the
posterior distribution of $(\beta,\Omega)$.  As in \cite{MR2036762},
we write
\[
\beta_i =
\begin{cases}
  (\beta_{i1}, \dots, \beta_{ii})' &\text{ if } i\le k,\\
  (\beta_{i1}, \dots, \beta_{ik})' &\text{ if } i> k,
\end{cases}
\]
and explicitly involve the latent factors in $F$.  Let $F_i$ be the
$n\times i$ matrix made up of the first $i$ columns of $F$, and write
$Y_i$ for the $i$-th column of $Y$ (in contrast to $y_t$, which is the
$t$-th row of $Y$).  The full conditionals for $F$, $\Omega$ and
$\beta$ are determined as follows.  First, the rows $f_t$ of $F$ are
conditionally independent given $(\beta,\Omega,Y)$ with
\begin{equation}
  \label{eq:fullcond:F}
  (f_t \,|\,\beta, \Omega,Y) \;\sim\; N_k\left((I_k  + \beta'\Omega^{-1}\beta)^{-1} \beta' \Omega^{-1} y_t   , ({I}_k + {\beta}'{\Omega}^{-1}{\beta})^{-1}\right)
\end{equation}
for $t = 1, \dots, n$.  Second, the uniquenesses
$\omega_1^2,\dots,\omega_m^2$ are conditionally independent given
$(\beta,F,Y)$ with
\begin{equation}
  \label{eq:fullcond:omega}
(\omega^2_i \,|\,  \beta,F,Y) \;\sim\; \mathit{IG}\left(\tfrac{1}{2}(\nu + T),\tfrac{1}{2}(\nu
s^2 + d_i)\right),
\end{equation}
where 
\[
d_i = ({ Y}_i - {F}_i{\beta}_{i}' )'({Y}_i - {F}_i{\beta}_{i}').
\]
Third, the rows of $\beta$ are conditionally independent given $(\Omega,
F, Y)$.  For $i = 1, \dots, k$, the conditional density of the
vector $\beta_i$ is proportional to
\begin{equation}
  \label{eq:fullcond:beta:1}
   \beta_{ii}^{k - i}\frac{1}{\det(C_i)}\exp\left\{-\frac{1}{2}({\beta}_i - { m}_i)' {C}_i^{-1}({\beta}_i - {
     m}_i)\right\}\mathbf{ 1}_{\{\beta_{ii}>0\}},
\end{equation}
where 
\[
{C}_i
= \left(\frac{1}{C_0}{ I}_i + \frac{1}{\omega_i^{2}}{ F}_i'{
    F}_i\right)^{-1}
\quad\text{and}\quad 
{m}_i = \frac{1}{\omega_i^{2}}{C}_i   {F}_i' { Y}_i .
\]
For $i = k+1, \dots, m$, the conditional distribution is
\begin{equation}
  \label{eq:fullcond:beta:2}
  ({ \beta}_i\,|\,  \Omega,F,Y) \; \;\sim\; N_k({ m}_i, { C}_i)
\end{equation}
with
\[
{C}_i
= \left(\frac{1}{C_0}{ I}_k + \frac{1}{\omega_i^{2}}{ F}'{
    F}\right)^{-1}
\quad\text{and}\quad 
{m}_i = \frac{1}{\omega_i^{2}}{C}_i   {F}' { Y}_i .
\]

The only full conditional that differs from those given in
\cite{MR2036762} is the one for $\beta_i$ with $i\le k$
from~(\ref{eq:fullcond:beta:1}).  To draw from this distribution, we
first sample from $(\beta_{ii}\,|\, \Omega,F,Y)$ and then from
$(\beta_{i1},\dots,\beta_{ii-1}\,|\,\beta_{ii},\Omega,F,Y)$.  The
latter distribution is a multivariate normal distribution.  The only
new challenge is thus the sampling from $(\beta_{ii}\,|\,
\Omega,F,Y)$, which has density proportional to
\[
\beta_{ii}^{k - i} e^{-\frac{(\beta_{ii} - a)^2}{2 b^2}}{\bf
  1}_{\{\beta_{ii} > 0\}}
\]
for constants $a \in \mathbb{R}$ and $b>0$ determined by
$(\Omega,F,Y)$.  After scaling $\beta_{ii}$ by $b$, the problem
reduces to generating draws from distributions with density in the class
\begin{equation}
  \label{eq:new:distribution}
  f(x\,|\, \alpha,\gamma) = \frac{1}{Z(\alpha,\gamma)} x^{\alpha-1}
    e^{-(x-\gamma)^2}, \quad x>0,
\end{equation}
where $\alpha>0$ and $\gamma\in\mathbb{R}$ are two parameters, and
$Z(\alpha,\gamma)$ is the normalizing constant.  In the present
context, integer values of $\alpha$ are of interest.  The densities
in~(\ref{eq:new:distribution}) are log-concave, and we use adaptive
rejection sampling \citep{gilks1992adaptive} as implemented in the R
package \texttt{ars} to generate from them.

\section{Numerical experiments}
\label{sec:numerics}

We illustrate the use of the two different priors, obtained
from~(\ref{eq:beta:prior:standard}) and~(\ref{Lprior}), respectively,
on a simulated dataset $Y$ that involves $m=15$ variables and is of
size $n=30$.  The data are drawn from the $k=3$ factor distribution
given by the following loading matrix and uniquenesses:
 \newcommand{\sss}{\scriptstyle }
\begin{equation*}
  \beta_0 = \bordermatrix{
    & \sss 1 & \sss 2 & \sss 3 \cr
    \sss 1 & 0.97 &  0    & 0    \cr 
    \sss 2 &  0.04 &  0.90 & 0    \cr 
    \sss 3 &  1.00 & -1.12 & 0.57 \cr 
    \sss 4 &  2.03 &  0.42 & 0.57 \cr 
    \sss 5 &  0.31 &  0.47 & 0.09 \cr 
    \sss 6 &  0.43 & -0.21 & -0.35\cr 
    \sss 7 &  0.75 &  0.31 & 0.68 \cr 
    \sss 8 &  0.45 & -0.48 & -1.50\cr 
    \sss 9 & -2.21 &  1.45 & 0.38 \cr 
    \sss 10 &  1.98 & -0.30 & 0.96 \cr 
    \sss 11 & -2.63 &  0.41 & 1.09 \cr 
    \sss 12 & -0.72 &  1.39 & 0.97 \cr 
    \sss 13 & -0.88 &  2.01 & -0.39\cr 
    \sss 14 & -0.53 &  0.04 & 0.59 \cr 
    \sss 15 & -0.95 &  1.39 & 0.37 \cr 
  }, \qquad  
  \text{diag}(\Omega_0) = \bordermatrix{
    &\cr
    \sss 1 & 0.17 \cr 
    \sss 2 &  0.05 \cr 
    \sss 3 &  0.02 \cr 
    \sss 4 &  0.02 \cr 
    \sss 5 &  0.05 \cr 
    \sss 6 &  0.06 \cr 
    \sss 7 &  0.04 \cr 
    \sss 8 &  0.67 \cr 
    \sss 9 & -0.04 \cr 
    \sss 10 & 0.21 \cr 
    \sss 11 & 0.10 \cr 
    \sss 12 & 0.09 \cr 
    \sss 13 & 0.21 \cr 
    \sss 14 & 0.51 \cr 
    \sss 15 & 0.03 \cr 
  }.
\end{equation*}
We create a second data matrix $Y^\pi$ by permuting the columns of $Y$
based on the permutation $\pi$ from Table \ref{tablepi}, i.e. the
$i$-th column of $Y$ becomes the $\pi(i)$-th column of $Y^{\pi}$.  For
Bayesian inference, we choose the hyperparameters as \cite{MR2036762},
that is, $C_0 = 1$, $\nu= 2.2$ and $s = \sqrt{0.1/2.2}$.  Via Gibbs
sampling, we draw from the posterior distributions for
the covariance matrix $\Sigma=\Omega+\beta\beta'$ for each data set,
focusing on the factor analysis models $k=3$, and $k=6$ factors.  The
Gibbs samplers are initialized at the respective maximum likelihood
estimates for $(\beta, \Omega)$.  After a burnin of $10, 000$
iterations, we ran each sampler for $300, 000$ iterations.

\begin{table}[t]
\centering
\caption{The permutation $\pi$ used to reorder simulated data.}
\label{tablepi}
\begin{tabular}{c|rrrrrrrrrrrrrrr}
  \hline
  \hline
$i$ &   1 &   2 &   3 &   4 &   5 &   6 &   7 &   8 &   9 &  10 &  11 &  12 &  13 &  14 &  15 \\ 
\hline
  $\pi(i)$ &  10 &   14 &  13 &  15 &  12 &   6 &   7 &  2 &  11 &   9 &   8 &   3 &   5 &   1 &   4 \\ 
   \hline
\end{tabular}
\end{table}

\begin{figure}[t]
\centering
      \begin{minipage}{.48 \linewidth}
              \includegraphics[width=\linewidth]{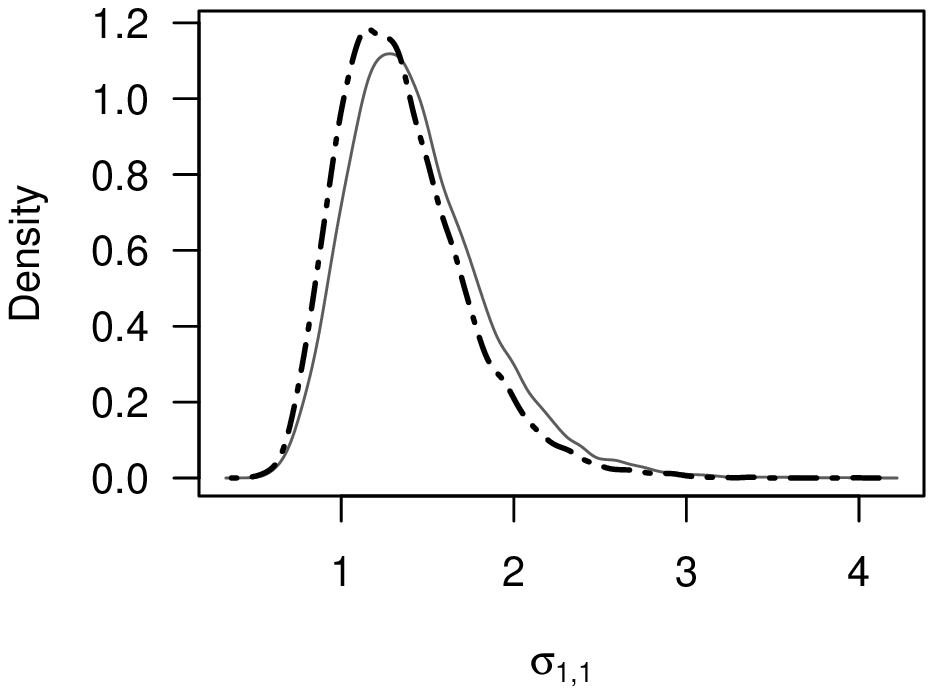}  
      \end{minipage} 
      \begin{minipage}{.48 \linewidth}
              \includegraphics[width=\linewidth]{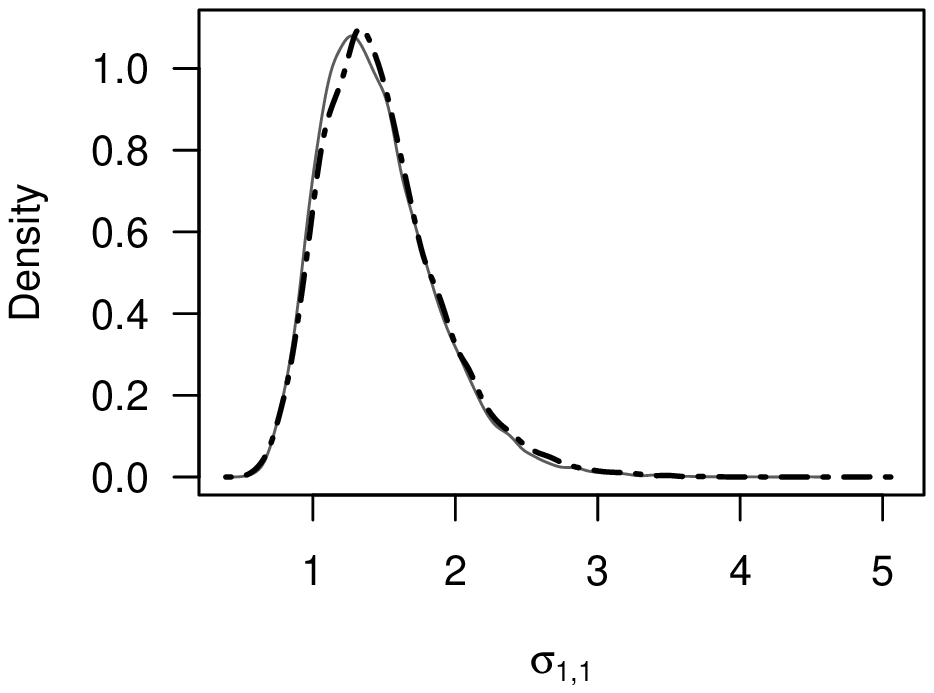}  
      \end{minipage}   
     
     \begin{minipage}{.48 \linewidth}
              \includegraphics[width=\linewidth]{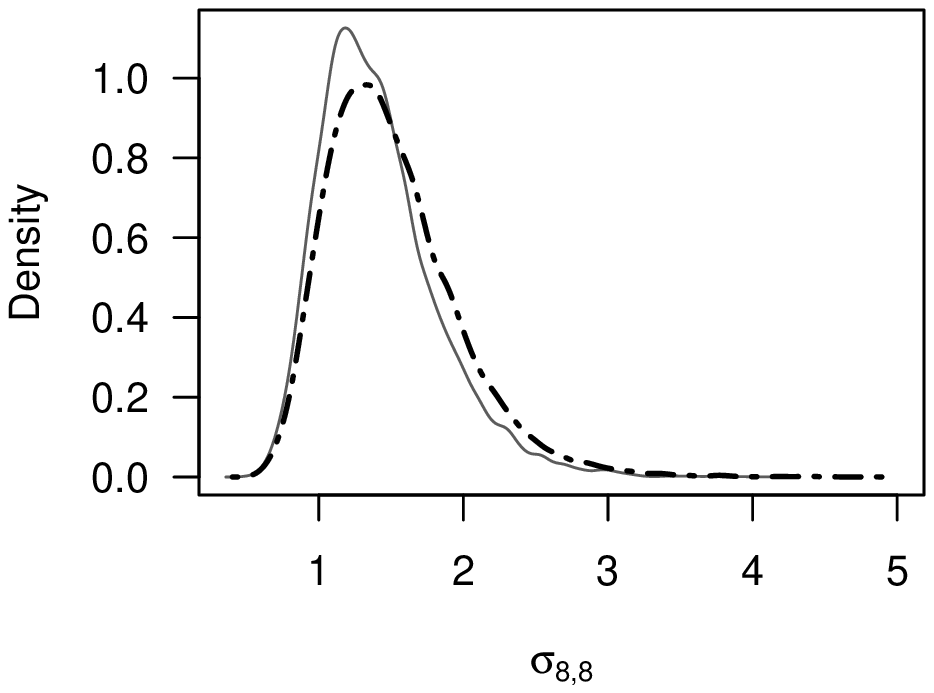}  
      \end{minipage}
 \begin{minipage}{.48 \linewidth}
              \includegraphics[width=\linewidth]{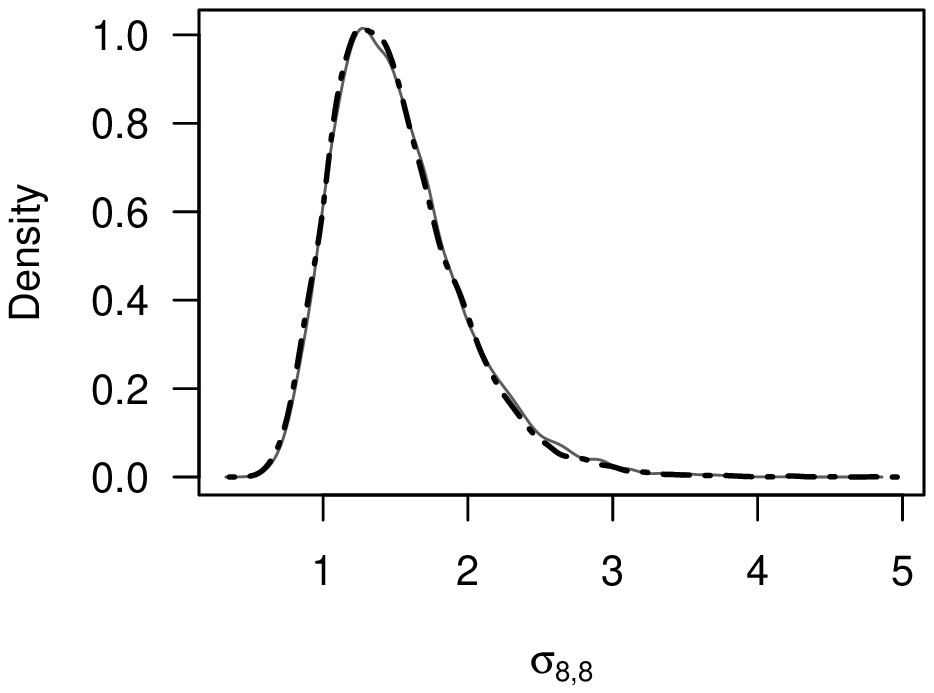}  
      \end{minipage}

      \begin{minipage}{.48 \linewidth}
              \includegraphics[width=\linewidth]{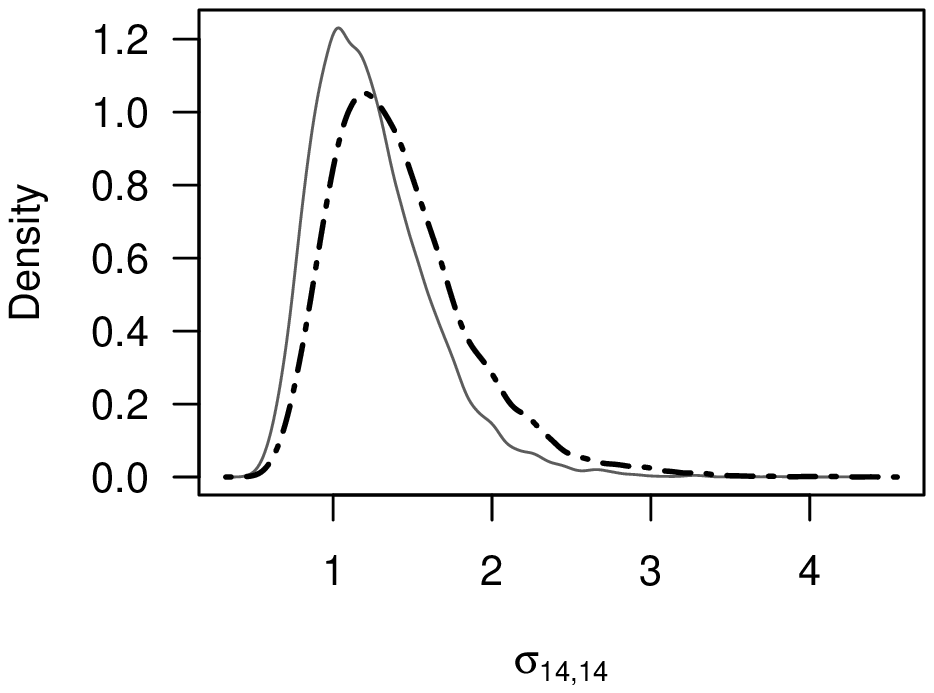}  
      \end{minipage}   
      \begin{minipage}{.48 \linewidth}
              \includegraphics[width=\linewidth]{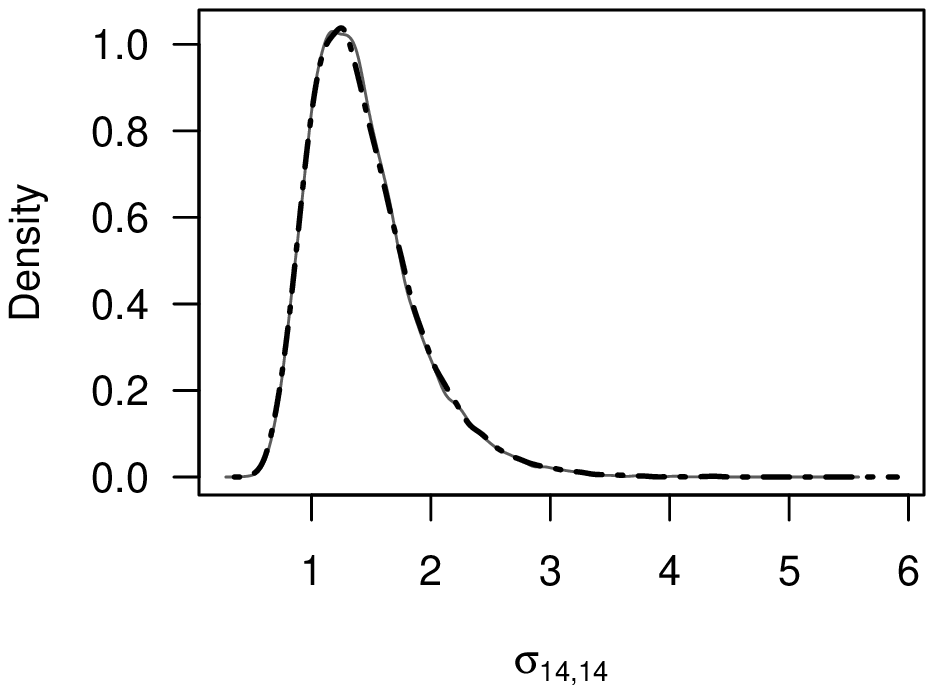}  
      \end{minipage}
      \caption{Posterior densities of $(\sigma_{ii}\,|\,Y)$, in black,
        and of $(\sigma_{\pi(i),\pi(i)}\,|\,Y^\pi)$, in grey, in
        factor analysis with $k=3$ factors, for $i=1,8,14$.
        The left column concerns the prior
        from~(\ref{eq:beta:prior:standard}), and the right column is
        based on the prior proposed in~(\ref{Lprior}).  }
\label{figurek3}
\end{figure}

\begin{figure}[t]
\centering
      \begin{minipage}{.48 \linewidth}
              \includegraphics[width=\linewidth]{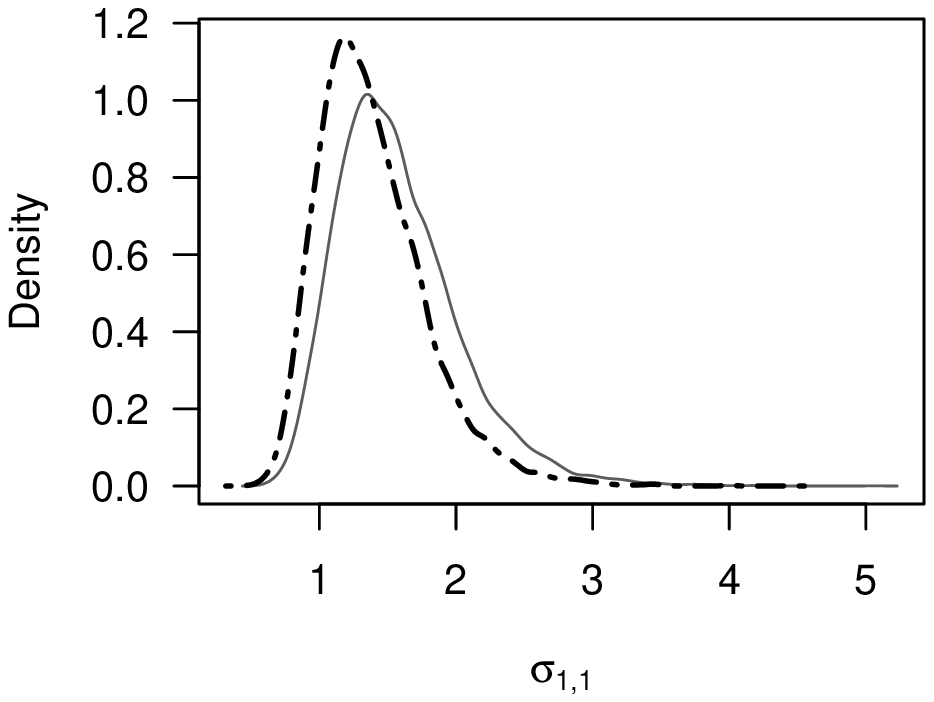}  
      \end{minipage} 
      \begin{minipage}{.48 \linewidth}
              \includegraphics[width=\linewidth]{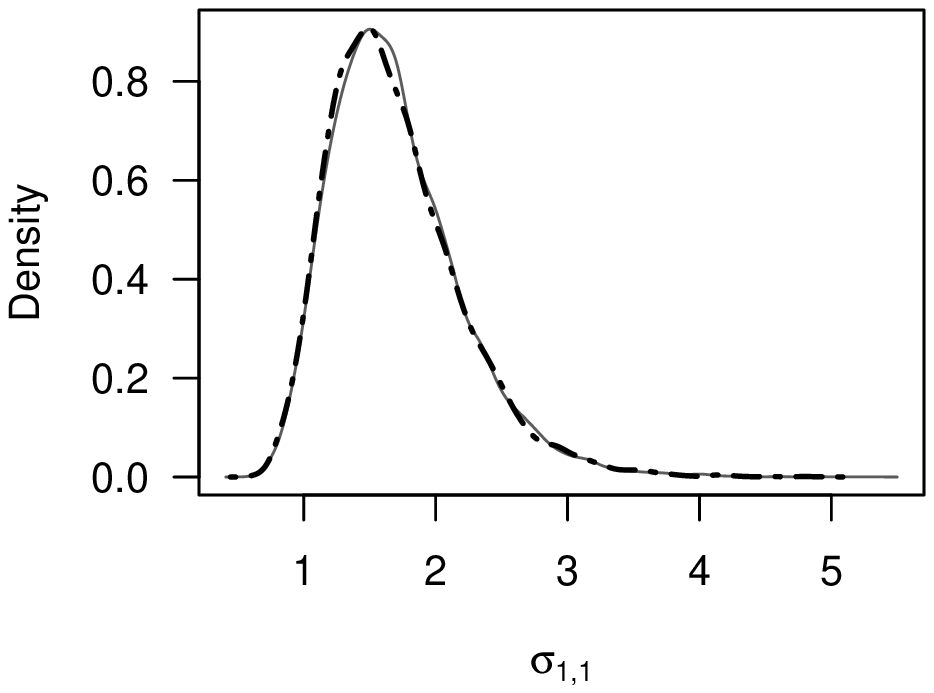}  
      \end{minipage}   
     
     \begin{minipage}{.48 \linewidth}
              \includegraphics[width=\linewidth]{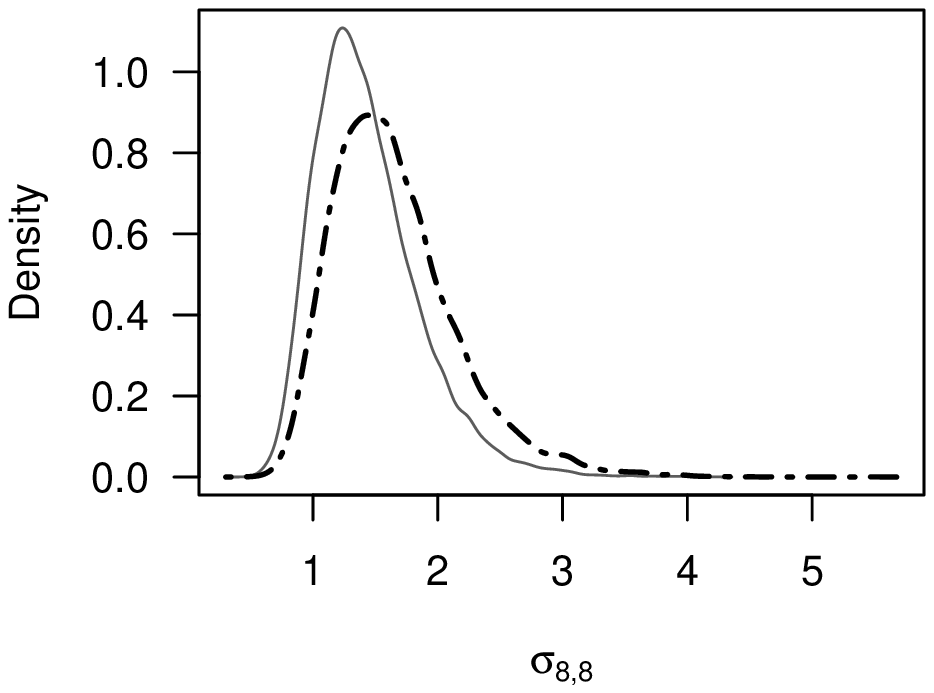}  
      \end{minipage}
 \begin{minipage}{.48 \linewidth}
              \includegraphics[width=\linewidth]{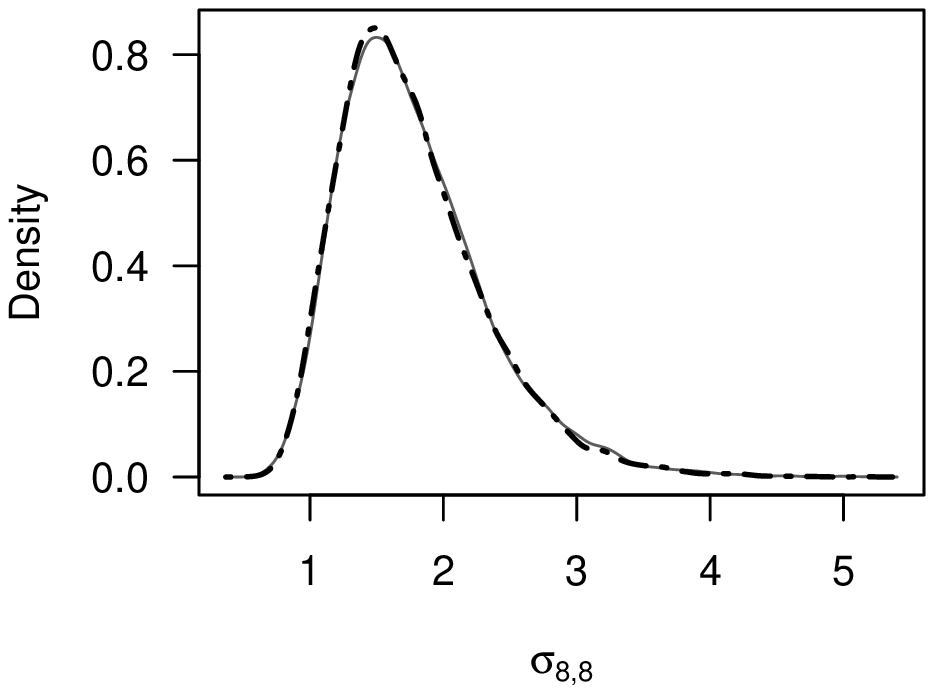}  
      \end{minipage}

      \begin{minipage}{.48 \linewidth}
              \includegraphics[width=\linewidth]{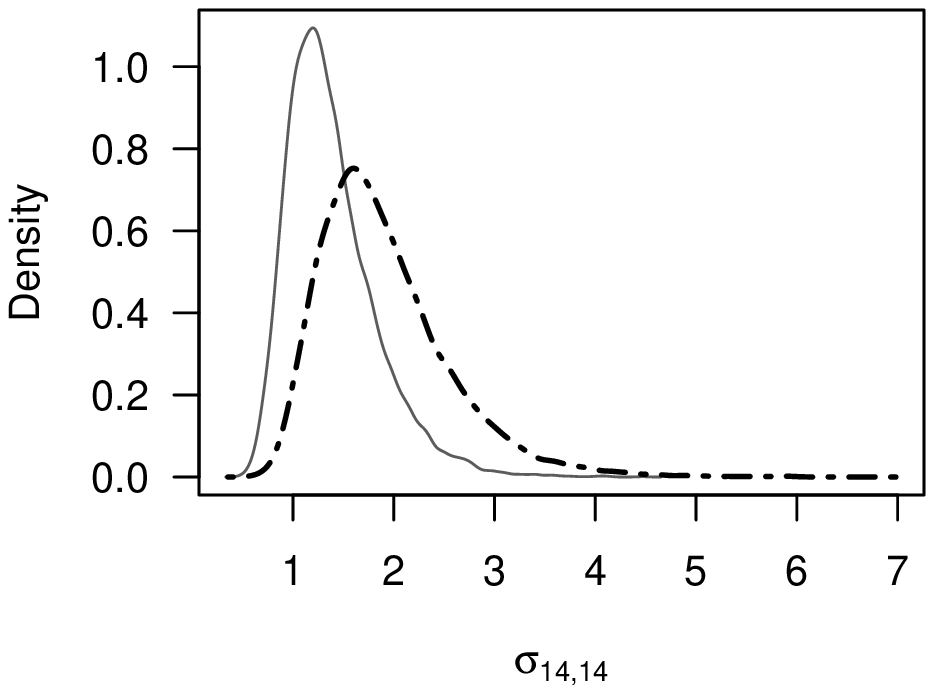}  
      \end{minipage}   
      \begin{minipage}{.48 \linewidth}
              \includegraphics[width=\linewidth]{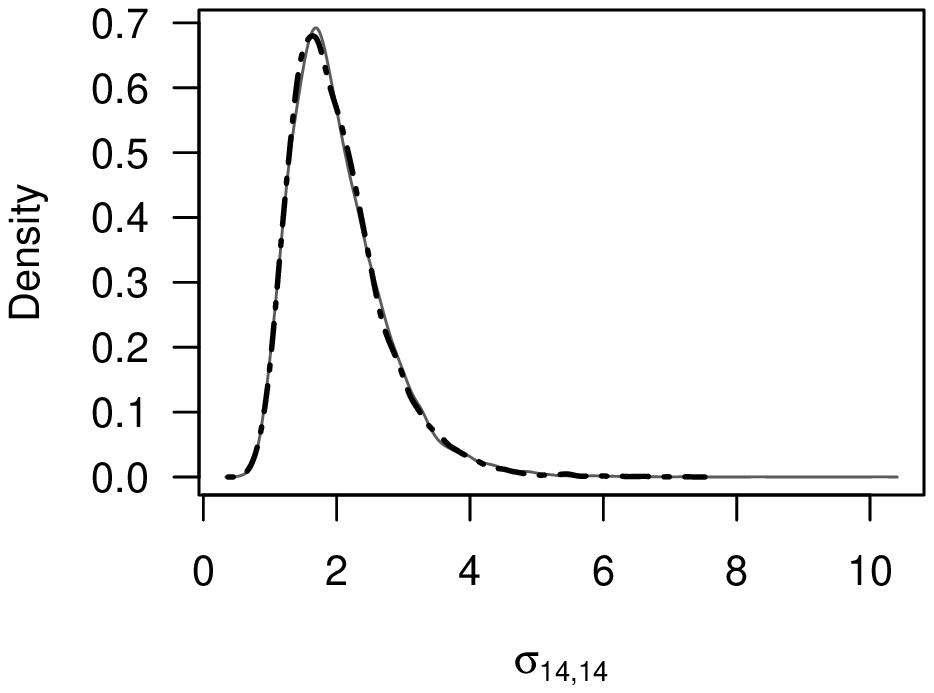}  
      \end{minipage}
\caption{Posterior densities of $(\sigma_{ii}\,|\,Y)$, in black, and
  of $(\sigma_{\pi(i),\pi(i)}\,|\,Y^\pi)$, in grey, in factor analysis
  with $k=6$ factors, for $i=1,8,14$.
  The left column concerns the
  prior from~(\ref{eq:beta:prior:standard}), and the right column is
  based on the prior proposed in~(\ref{Lprior}).  }
\label{figurek6}
\end{figure}

Figures~\ref{figurek3} and~\ref{figurek6} show kernel density
estimates of the posterior densities of selected variances.  More
precisely, we compare the densities of $(\sigma_{ii}\,|\,Y)$ and
$(\sigma_{\pi(i),\pi(i)}\,|\,Y^\pi)$ for $i=1,8,14$.  Under our
proposed prior from~(\ref{Lprior}), the two posterior densities are
the same.  Indeed, the plots in the right hand columns of
Figures~\ref{figurek3} and~\ref{figurek6} show only minor
discrepancies due to Monte Carlo error.  The `standard prior'
from~(\ref{eq:beta:prior:standard}), however, results in visible
differences that are more pronounced for $k=6$, which is not
surprising as larger differences are possible among the degrees of
freedom of the chi-square prior for $(\beta\beta')_{ii}/C_0$;
recall~(\ref{eq:betabeta:ii:chisquare}).  Note that the observed
shifts in the posterior distributions under the `standard prior' are
explained by the different chi-square degrees of freedom.

\section{Conclusion}
\label{sec:conclusion}

This paper proposes a prior distribution for the loading matrix in
factor analysis.  The proposal allows for computation with an
identifiable lower triangular loading matrix $\beta$ all the while
having the associated covariance matrix invariant under permutation of
the variables at hand.  The prior is intended as a possible default
when there is no reason to impose dependence among loadings or to
treat the loadings of different variables differently.  Concerning
possible departures from our default scenario, we remark that the
software of \cite{mcmcpack} also allows one to impose patterns of
zeros in the loading matrix $\beta$.  As mentioned earlier, the latter
situation is sometimes termed confirmatory factor analysis.  The
identifiability issues we addressed need not arise in that case as
orthogonal transformations will generally not preserve prescribed
zeros in the loading matrix.

Sampling from the posterior distribution resulting from the prior we
proposed is largely the same as for the `standard prior' that has been
used by several authors including 
\cite{geweke:1996} and \cite{MR2036762}.  The key difference is the
need to sample from distributions in the class specified
by~(\ref{eq:new:distribution}).  These distributions also appear in
the realm of multivariate $t$-distributions
\citep{finegold:aoas,finegold:ba}, although a square-root
transformation is necessary to match the setup there.  It thus seems
worthwhile to develop an efficient sampler targeting precisely this
family of distributions, which is a problem we are working on.

Finally, we emphasize that our proposal rests 
in an important way on
the fact that we derived it from a spherical joint normal distribution
for the loading matrix, namely, $\beta\sim N_{m\times
  k}(0,C_0I_m\otimes I_k)$.  Departures from this situation, even
merely including a non-zero mean for this matrix normal distribution,
seem to lead to a considerably more difficult scenario.

  \section*{Acknowledgments}

  This work was supported by the U.S.~National Science Foundation
  (DMS-1305154) and by the University of Washington Royalty Research
  Fund.

\bibliographystyle{ba}
\bibliography{fact_anal_bib}

\end{document}